\documentstyle[twocolumn,aps]{revtex}

\begin{document}
\author{Eric L. Bolda, Sze M. Tan and Dan F. Walls}

\address{Department of Physics, University of Auckland, Private Bag 
92019,\\ 
Auckland, New Zealand.}

\title{Reconstruction of the joint state of a two-mode Bose-Einstein condensate}

\date{\today}
\draft

\maketitle
  
\begin{abstract}
We propose a scheme to reconstruct the state of a two-mode 
Bose-Einstein condensate, with a given total number of atoms, using an 
atom interferometer that requires beam splitter, phase shift and 
non-ideal atom counting operations.  The density matrix in the 
number-state basis can be computed directly from the probabilities of 
different counts for various phase shifts between the original modes, 
unless the beamsplitter is exactly balanced.  Simulated noisy data 
from a two-mode coherent state is produced and the state is 
reconstructed, for 49 atoms.  The error can be estimated from the 
singular values of the transformation matrix between state and 
probability data. 
\end{abstract}



In quantum physics, a system can be completely described by its 
quantum state.  On the other hand, in a given experiment one can only 
measure the probability of a quantity, and the process of doing so 
usually destroys any knowledge of the complementary aspect of that 
quantity.  Thus to reveal the whole state a variety of measurements 
must be made on an ensemble of identically prepared systems.  Even 
when the systems are not identical one can recover a mixture of 
quantum states in the form of the density matrix.  Optical homodyne 
tomography (OHT) is one such method of revealing the density matrix, 
tailored to determine the state of an electromagnetic field 
mode\cite{Leonhardt95}.  In that case the availability of coherent 
states from lasers allows the measurement of the probability of a 
quadrature value, for a given phase difference between the coherent 
state and the signal state.  It also lends itself to a simple 
geometric interpretation: the measured distributions are the 
``shadows'' of rotations of the Wigner function.  Recent work has 
shown that it may be easier numerically to reconstruct the density 
matrix in the number-state basis directly from the distributions, 
without first going through the Wigner function 
\cite{OHTrev}.

In the field of atom optics, breakthroughs in the evaporative cooling of 
atomic vapors have 
resulted in observation of Bose-Einstein condensates (BEC)  
\cite{BECexpt}.  We may view these as the analog 
of a single-mode optical field, except that the field now has a finite 
mass and a quartic interaction term.  The ``semiclassical'' aspects of 
BEC such as collective excitations and condensate shape are now a 
subject of thorough experimental investigation \cite{JinMewes}.  
However, perhaps the most interesting aspects are due to the quantum 
states themselves.  We might expect unusual quantum states for several 
reasons.  One is, the collisions are analogous to nonlinear optical 
susceptibilities which are known to produce squeezing 
\cite{squeezing,Dunningham97a}.  Second, an entangled state with a 
random relative phase can emerge between two condensates, when they 
produce spatial interference fringes in the atom counts 
\cite{spatialint}.  Also, in a closed system, quantum 
superpositions of different total numbers are not permitted (although 
mixtures of different numbers are).  For this reason, the OHT method 
cannot be applied to these BEC since coherent states are such 
superpositions of number states.  A recent experiment \cite{Mewes96a} 
has demonstrated coherence within a single output pulse of atoms; 
however, the phase varies from one pulse to the next, and the exact 
nature of the output state is not yet known.  Although theoretical 
work on atom lasers suggests that coherent atom fields will eventually 
be available \cite{bosers}, we instead take a 
different tack and reconstruct the density matrix from a two-mode 
system with fixed atom number.

We describe an atom interferometer, and a set of measurements 
sufficient to reconstruct the state.  Using an angular momentum 
formalism suitable for the two-mode states, the reconstruction process 
amounts to the solution of a linear algebra problem.  This is 
accomplished with standard Fourier transform and least-squares 
methods.  Finally we reconstruct a two-mode coherent state from 
simulated noisy data, and compare the actual error with an analytical 
estimate.


We consider a fixed number of atoms which have been cooled to below 
the critical temperature for BEC. The atoms are prepared in the 
two-mode quantum state
\begin{equation}
 \rho = \sum_{a} p_{a} |\psi_{a} \rangle \langle \psi_{a} |
\end{equation}  
\begin{equation} |\psi_{a} 
\rangle = \sum_{n=0}^{N}  c_{n,a} |N - n \rangle_{1} \otimes |n \rangle_{2}  
\end{equation} 
that is to be measured.  The modes may correspond to  {\it e.\  g.\ } 
different hyperfine states; such a BEC has just been produced in an 
experiment \cite{Myatt97}.  One of the modes has its phase shifted 
relative to the other by $\phi$, that can be varied. 
Then an operation corresponding to that of a beam splitter is applied to 
the two modes.  The number of atoms in one of the output modes is 
counted with the detector (see Fig.\ \ref{BEMZI}).  If the initial 
state preparation is considered to be a generalization of a beam 
splitter (which entangles a product state), the interferometer 
resembles a Mach-Zehnder arrangement. To collect data 
for tomography, for each of $K$ phase shifts, we repeat the experiment 
$\tau$ times, resulting in probability distributions of counts at each 
value of $\phi$.


At this stage we are making several idealisations.  We neglect the 
collisions between the atoms, especially at the beam splitter where 
this would result in mixing the two modes.  The beamsplitter is 
assumed to be lossless, with transmission $\cos^{2}\theta$.  The 
action of the phase shifter and beamsplitter together on the density 
matrix $\rho$ for the two modes is given by
\begin{equation}
\rho_{out} = U(\theta, \phi)^{\dag} \rho U(\theta, \phi)
\end{equation}
with
\begin{equation}
U(\theta, \phi) = \exp[i \theta (a_{1}^{\dag} a_{2} e^{i \phi} + 
a_{2}^{\dag} a_{1} e^{i\phi})].
\end{equation}

Finally we assume the detector has unit efficiency.  This assumption 
is not crucial, as we can accumulate the same set of data when the 
efficiency is less than unity by counting atoms in both modes and 
keeping only the data for which the sum of counts was $N$.

One possible realisation of the interferometer is an extension of the 
recent output coupler built by the MIT group {\it et al} 
\cite{Mewes96a}.  After the condensate is prepared in a superposition 
of $|g m_{F} \rangle$ and $|g m_{F} +2 \rangle$ hyperfine states, the 
atoms are dropped out of the trap.  The operation $U(\theta, \phi)$ is 
accomplished via two off-resonance optical pulses, connecting the 
states in a Raman transition, which also introduces a relative phase 
shift due to the phase between the pulses.  
Finally an on-resonance rf pulse transfers atoms from the untrapped 
$|g m_{F} \rangle$ state to the trapped $|g^{\prime} m_{F} \rangle$ 
state; the separate groups of atoms can then be counted.  

To simplify the notation and the calculation of 
matrix elements, we use the formal equivalence between the 
algebra  for two harmonic oscillators and that for angular momentum 
\cite{Sakurai85}.  We write 
the state
\begin{equation}
|n \rangle_{1} \otimes | N - n \rangle_{2} = |j+m \rangle_{1} \otimes | 
j - m \rangle_{2}  
\: \mbox{ as} \: |m \rangle
\end{equation}
where $j = N/2$ and $m = n - j$.  The $2j+1$ states $|m \rangle$ have all the 
properties of the eigenstates of $J^{2}$ and $J_{z}$ and important 
operators are
\begin{eqnarray}
J_{+} & = & a_{1}^{\dag} a_{2} \\
J_{-} & = & a_{2}^{\dag} a_{1} \\
J_{z} & =  &\frac{1}{2} \left(a_{1}^{\dag} a_{1} - a_{2}^{\dag} a_{2} 
\right).
\end{eqnarray}
The combined effect of the beamsplitter and phase shift is seen to be a rotation 
by $-2 \theta$ about an axis $\bf{\hat{n}}_{\phi} = \bf{\hat{x}} \cos \phi - 
\bf{\hat{y}} \sin \phi$,
\begin{eqnarray}
U(\theta, \phi) & = & \exp \left[i \theta (J_{+} e^{i \phi} + J_{-} e^{-i 
\phi})\right] \nonumber \\
 & = & \exp \left[i 2 \theta \mathbf{J} \cdot \bf{\hat{n}}_{\phi} 
 \right].
\end{eqnarray}
An explicit expression for the number-basis matrix elements of the rotation is
\begin{eqnarray}
{\cal D}_{lm}^{(j)}  & \equiv & \langle l| U(\theta, \phi) | m \rangle 
\nonumber \\
 & = & \sum_{k} 
 \frac{\sqrt{(j+m)!(j-m)!(j+l)!(j-l)!}}{k!(j+m-k)!(j-k-l)!(k+l-m)!} 
 \nonumber \\ 
 & \times & (\cos \theta)^{2j-2k+m-l} (i \sin \theta)^{2k+l-m} e^{i(l-m)\phi}.
 \label{Dnumber}
\end{eqnarray}
where the sum is over all values for which the arguments of the 
factorials are nonnegative.

 The probability of $m$ counts at the detector, for a 
phase shift setting of $\phi$ is
\begin{equation}
P_{m}(\phi) = \langle m |U(\theta, \phi)^{\dag} \rho_{in} U(\theta, 
\phi)| m \rangle
\end{equation}
or in the number-state basis
\begin{equation}
P_{m}(\phi) = \sum_{l,l^{\prime}} T_{m,\phi l,l\prime} 
\rho_{ll\prime},
\label{Pm}  
\end{equation}
where
\begin{equation}
T_{m,\phi l,l{\prime}} = {\cal D}_{lm}^{(j) *}(\theta, \phi) {\cal 
D}_{l\prime m}^{(j)}(\theta, \phi).
\label{Tnumber} 
\end{equation}  
This last equation shows that $T$ is simply a linear operator taking 
the density matrix (an $(N+1)^{2}$ element complex vector) to the probability 
data (an $(N+1) K$ element real vector); we wish to invert this 
operator to reconstruct the density matrix.  This is formally done by 
defining the positive semidefinite, hermitian $(N+1)^{2}\times (N+1)^{2}$ matrix
\begin{equation}
M = T^{\dag} T,
\end{equation}
where $T^{\dag}$ means the conjugate transpose of the matrix $T$, and computing
\begin{equation}
\rho = M^{-1} T^{\dag} P,
\end{equation}
provided $M$ is nonsingular.  This is actually  
the best solution in the least squares sense \cite{Bjorck96}. 

For the value $\theta = \frac{\pi}{4}$, corresponding to a balanced 
beamsplitter, the matrix $M$ is singular.  Since this might be 
somewhat surprising we prove this result, and give an 
example of two states which will produce identical probability data. First we 
note that if there exists a vector $v$ such that $M v = 0$, this 
implies $T v = 0$.  Making the {\em Ansatz}
\begin{eqnarray}
v = f(J_{z}) \\
f(-q) = -f(q),
\end{eqnarray}
and using the formula for $T$ in the number state basis Eq. 
(\ref{Tnumber}) we find
\begin{equation}
Tv = \sum_{q=-j}^{j}|{\cal D}_{l,q}^{(j)}(\frac{\pi}{4}, \phi)|^{2} f(q).
\label{Tvpi4}
\end{equation}
Then, using the fact from Eq. (\ref{Dnumber}) that
\begin{equation} 
|{\cal D}_{l,q}^{(j)}(\frac{\pi}{4}, \phi)| =  |{\cal 
D}_{l,-q}^{(j)}(\frac{\pi}{4}, \phi)|,
\end{equation}
yields zero for the sum of an odd function in Eq. (\ref{Tvpi4}), 
for any values of $l$ and $\phi$. This 
proves that the matrix $M$ is singular for 
$\theta = \frac{\pi}{4}$. To illustrate this physically, we choose the pure states
\begin{eqnarray}
|\psi_{1} \rangle = |\frac{N}{2} + m \rangle_{1} \otimes |\frac{N}{2} - m 
\rangle_{2} \\
|\psi_{2} \rangle = |\frac{N}{2} - m \rangle_{1} \otimes |\frac{N}{2} + m 
\rangle_{2}
\end{eqnarray}
having the respective density matrices
\begin{eqnarray}
\rho_{1} & = & |m \rangle \langle m| \\
\rho_{2} & = & |-m \rangle \langle -m|.
\end{eqnarray}
Then the difference vector of the density matrices is in the 
nullspace of T:
\begin{equation}
\rho_{2} - \rho_{1} = f(J_{z})  \: \mbox{where} \:  f(x) = 
\left\{ 
 \begin{array}{ll}
   1 & \mbox{ if $ x=m$} \\
  -1 & \mbox{if $x=-m$}  \\
   0 & \mbox{otherwise} 
 \end{array}
\right. 
\end{equation}
so that $T \rho_{1} = T \rho_{2}$.
Clearly the balanced scheme cannot distinguish these two states because there 
is an ambiguity about which number of atoms belongs to which mode.
This difficulty does not arise in OHT because in that case one of the 
inputs is already known to be a coherent state.  
Of course, varying $\theta$
will enable these two situations to be distinguished, but we would 
like to keep the experimental set up the same for all of the data 
collection, so we require $\theta \neq \frac{\pi}{4}$.


The problem of solving Eq. (\ref{Pm}) is considerably simplified by 
taking the Fourier series of the probability data,
\begin{eqnarray}
F_{m}(r)  & = & \frac{1}{2 \pi} \int_{0}^{2 \pi} P_{m}(\phi) e^{i r \phi} 
 \, d \phi  \label{fmr}  \\
 & = & \sum_{l = r-j}^{j} T_{m0,l, l-r} \, \rho_{l,l-r}.
\end{eqnarray}
(In practice, this is approximated by the discrete Fourier 
transform on a set of data for many $\phi$ equally distributed about 
the circle.)  By taking succesive values of $r$ from $0$ to $2j$, each 
diagonal of the density matrix can be solved for independently.  Since 
all the operations are linear the calculations are conveniently done in 
MATLAB. 

We will see below that for elements far from the main diagonal, the 
data can be insufficient to accurately reconstruct them. Since 
the trace of $\rho^{2}$ is less than unity, each norm of a diagonal is 
also less than one.  Taking this prior information into account, we 
can reduce the norm of each diagonal by using Tikhonov 
regularization \cite{Hanson94}.

We present a simulation of the probability data and reconstruct the 
density matrix from it, for the pure state
\begin{equation}
| \psi \rangle = \sum_{n=0}^{N} \sqrt{\frac{N!}{(N-n)!n!}} 
\sin^{N-n} \vartheta
\cos^{n} \vartheta \; e^{i \varphi} |n \rangle_{1} \otimes |N-n \rangle_{2}.
\label{state}.
\end{equation}
This state is the projection of coherent states for modes $1$ and $2$ 
onto the subspace with fixed total number $N$ \cite{Molmer97}; it can 
be produced by an rf pulse as in \cite{Mewes96a}.
The parameters chosen were $N = 49$, $\vartheta = 0.54$, $\varphi = 0.13$.
To simulate the noise associated with calculating 
probabilities from histograms, we add to each probability a term
\begin{equation}
\delta P_{m}(\phi) = \sqrt{\frac{P_{m}(\phi)}{\tau}} g(P_{m}(\phi)),
\label{pnoise}
\end{equation}
where $\tau$ is the number of trials for that value of phase and $g$ 
is a Gaussian distribution with zero mean and unit variance.  In the 
example below we take $\tau = 2000$ and $K= 180$, {\it i.\ e.\ }  2000 
trials every 2 degrees.  In Fig.  \ref{probm} we plot probability of 
$m$ counts at each phase shift.  The original and reconstructed 
density matrices are compared in Fig.  \ref{rhofig}.  The 
reconstruction is most accurate (in both magnitude and phase) for the 
elements close to the main diagonal.

The error in the solution to the least squares problem can be found 
using the singular value decomposition of each matrix $T$ 
\cite{Bjorck96}.  Specifically, if the norm of the error in $F(r)$ is
$\| \delta F(r) \|$ then the norm of the difference between the 
original and reconstructed $r$-th diagonal is
\begin{equation}
\| \delta \rho_{r} \|  \equiv \sqrt{\sum_{i=-j}^{j-r} |\rho_{i,i+r}|^{2}}
\leq \frac{\| \delta F(r) \|}{\min \sigma},
\end{equation}
where $\sigma$ is an eigenvalue of $\sqrt{T^{\dag}T}$.  Using the 
definition Eq. (25) to transform the variances $\delta P_{m}$, and 
the fact that probabilities for a given phase must sum to one, we find
\begin{equation}
\| \delta F(r) \| = \frac{1}{\sqrt{K \tau}}.
\end{equation}
Thus our upper bound is independent of the actual data, depending
only on $N$, $r$ and $\theta$.  In Fig. \ref{errfig} we plot the 
actual and estimated norm of the error, divided by $\sqrt{N-r+1}$ so 
that it is per element, as a function of the distance $r$ from the main 
diagonal.  We have not been 
able to find a simple formula for the eigenvalues $\sigma = \sigma(N, 
r,\theta)$, but the general trend is for smaller minimum values as $r$ 
approaches $N$. The most difficult states to reconstruct will be
highly entangled ones, which depend on the high-frequency components 
of the probability.  Provided one has some idea 
of which diagonals are needed this analysis is useful in choosing the 
optimal transmission.

A different type of error might occur if the number of atoms 
originally trapped is not known precisely, or (equivalently) if losses 
occur between the preparation of the original state and the detection 
of atoms.  We are unable to quantify such errors, since states of 
fixed number are assumed.  For mixed states with narrowly 
distributed total numbesr, but no fine-scale oscillations in 
density matrix elements, we can expect the above algorithm will 
produce a qualitatively correct picture.

To conclude, we have proposed an atom interferometer for 
measuring the quantum state of a two-mode BEC.  An algorithm for 
calculating the density matrix elements was applied to simulated 
probability data and successfully reconstructed a sample state.  The 
errors can be estimated before data collection, and may be minimized 
by choosing an appropriate beamsplitter transmission if some 
qualitative features about the state are known.

This research was made possible by the Marsden Fund of the Royal 
Society of New Zealand and The University of 
Auckland Research Fund.  Frequent discussions with Tony Wong and Michael Jack 
provided many insights.  We would like to thank M.-O. Mewes 
for his detailed suggestions regarding the proposed interferometer.

\bibliographystyle{prsty}

\begin{figure}
\caption{Schematic of atom interferometer needed for state 
reconstruction. One of the modes acquires a phase shift $\phi$ before 
they both undergo a beamsplitter operation (BS). An atom counter (D) is 
placed at one or both outputs.}
\label{BEMZI}
\end{figure}

\begin{figure}
\caption{Probability as a function of number of atom counts at one 
port and relative phase shift between modes.  Computed from Eq. 
(\ref{Pm}) with simulated noise Eq. (\ref{pnoise}). }
\label{probm}
\end{figure}

\begin{figure}
\caption{Density matrix of {\em (a)} the original state and {\em (b)} 
the reconstructed state, in the number basis.  Magnitude is height 
and phase is color.} 
\label{rhofig}
\end{figure}

\begin{figure}
\caption{{\em Data points}: Norm of the difference, per element, in the original and 
reconstructed density matrices, as a function of distance from the 
main diagonal.  {\em Solid curve}: Estimated norm of the difference, 
per element, 
based on the singular values of the transformation matrices.}
\label{errfig}
\end{figure}

\end{document}